\documentclass[preprint,12pt]{elsarticle}

\usepackage{amsmath,amsfonts,nccmath}
\usepackage{algorithmic}
\usepackage[linesnumbered,ruled,noend]{algorithm2e}

\usepackage{textcomp}
\usepackage{stfloats}
\usepackage{verbatim}
\usepackage{makecell}
\usepackage{graphicx}
\usepackage{enumitem}
\usepackage{multirow}
\usepackage{soul}
\usepackage{xcolor}
\usepackage{color}
\usepackage{tikz}
\usepackage{amsmath}
\usepackage{url}
\usepackage{booktabs,threeparttable}
\usepackage{comment}
\usepackage{cuted}
\usepackage{pifont}
\usepackage{caption}
\usepackage{subcaption}
\usepackage{color, colortbl}
\usepackage{amsthm}
\usepackage{soul}
\usepackage{marvosym}

\definecolor{Gray}{gray}{0.85}
\definecolor{verylightgray}{rgb}{.97,.97,.97}
\definecolor{mygreen}{RGB}{24,141,31}
\definecolor{myred}{RGB}{142,0,8}
\definecolor{mypurple}{RGB}{107,29,111}

\usepackage{amssymb}
\usepackage{adjustbox}
\usepackage{hyperref}
\hypersetup{
    colorlinks=true,
    linkcolor=mygreen,
    filecolor=mygreen,      
    urlcolor=mygreen,
    citecolor=mypurple,
}

\begin{document}

\begin{frontmatter}



\title{Large Language Models for Blockchain Security:\\ A Systematic Literature Review}


\author[inst1]{Zheyuan He}

\affiliation[inst1]{organization={University of Electronic Science and Technology of China},
            country={China}}

\author[inst2]{Zihao Li}

\affiliation[inst2]{organization={The Hong Kong Polytechnic University},
            country={China}}

\author[inst1]{Sen Yang}

\author[inst3]{He Ye}

\affiliation[inst3]{organization={University College London},
            country={United Kingdom}}

\author[inst1]{Ao Qiao}

\author[inst1]{Xiaosong Zhang}

\author[inst1]{Ting Chen}

\author[inst2]{Xiapu Luo}

\begin{abstract}
Large Language Models (LLMs) have emerged as powerful tools across various domains within cyber security. Notably,
recent studies are increasingly exploring LLMs applied to the context of blockchain security (BS).
However, there remains a gap in a comprehensive understanding regarding the full scope of applications, impacts, and potential constraints of LLMs on blockchain security.
To fill this gap, we undertake a literature review focusing on the studies that apply LLMs in blockchain security (LLM4BS).

Our study aims to comprehensively analyze and understand existing research, and elucidate how LLMs contribute to enhancing the security of blockchain systems.
Through a thorough examination of existing literature, we delve into the integration of LLMs into various aspects of blockchain security. 
We explore the mechanisms through which LLMs can bolster blockchain security, including their applications in smart contract auditing, transaction anomaly detection, vulnerability repair, program analysis of smart contracts, and serving as participants in the cryptocurrency community.
Furthermore, we assess the challenges and limitations associated with leveraging LLMs for enhancing blockchain security, considering factors such as scalability, privacy concerns, and ethical concerns. 
Our thorough review sheds light on the opportunities and potential risks of tasks on LLM4BS, providing valuable insights for researchers, practitioners, and policymakers alike.
\end{abstract}


\end{frontmatter}


\section{Introduction}
\label{sec_intro}

As the digital era advances, the confluence of artificial intelligence with blockchain technology emerges as a groundbreaking development, particularly at the juncture where Large Language Models (LLMs)~\cite{kasneci2023chatgpt, greshake2023not,shen2024hugginggpt,yao2024survey,wu2023brief,ma2023scope} intersect with the ever-evolving domain of blockchain security~\cite{chang2023survey,liu2024your,wu2023brief,ma2023scope,zhang2023machine,wang2023financial,zhang2024rapid}. 
LLMs have risen to the forefront of blockchain security~\cite{zhang2019security,motlagh2024large}, 
showcasing profound capabilities in text generation and comprehension~\cite{sun2024gptscan,Hou2023LargeLM,ma2023scope,jin2023inferfix}), especially in source code analysis.
These abilities mirror human-like proficiency~\cite{ji2024beavertails,liu2023summary}. 
This transformative impact is attributable to their expansive datasets, sophisticated architectures, and the deep neural networks that underpin their operational frameworks~\cite{kasneci2023chatgpt,hu2023large,liu2023summary,deng2023pentestgpt}.

The robustness of LLMs in discerning and synthesizing complex patterns within data positions them as invaluable assets in enhancing the security measures within blockchain 
systems~\cite{li2020survey,luo2023bc4llm,shen2024hugginggpt,yao2024survey,azad2024machine, xu2024large, nguyen2024generative, mboma2023assessing, mboma2024integrating}.
Concretely, the granular analysis of smart contracts~\cite{sun2024gptscan}, the meticulous scrutiny of transactions~\cite{gai2023blockchain}, and automatic code (resp. text) generation~\cite{storhaug2023efficient} are among the critical tasks that LLMs are adept at performing with remarkable efficacy~\cite{Hou2023LargeLM,Zhao2023ASO,zhang2023machine}.

However, integrating these cognitive powerhouses into blockchain security is met with an array of challenges that beckon for consideration.
Navigating the intricate dynamics of ever-advancing cybersecurity threats and addressing the ethical concerns that accompany AI deployment make this trajectory as demanding as it is promising.
Despite the progress, there is still a lack of comprehensive work depicting the current application status and future development prospects of LLM in blockchain security (BS).

\begin{table}[t!]
\caption{Table of LLM4BS studies}
\label{tab_works}
\resizebox{\textwidth}{!}{%
\begin{tabular}{@{}lcl@{}}
\toprule
\textbf{Domains}     & \textbf{Amounts}    & \textbf{Publications} \\ \midrule
Smart Contract Auditing        & 22           & \begin{tabular}[c]{@{}l@{}}SMARTINV~\cite{wang2024smartinv}, GPTScan~\cite{sun2024gptscan}, David et al.~\cite{david2023you},\\ Karanjai et al.~\cite{karanjai2023smarter},  ContractArmor~\cite{sonmez2024contractarmor},  Ortu et al.~\cite{ortu4530467identifying},\\ ASSBert~\cite{sun2023assbert}, PSCVFinder~\cite{yu2023pscvfinder}, LLM4Vuln~\cite{sun2024llm4vuln},\\ TrustLLMf~\cite{ma2024combining}, AuditGPT~\cite{xia2024auditgpt}, PropertyGPT~\cite{liu2024propertygpt} \\
Chen et al.~\cite{chen2023chatgpt}, Jain et al.~\cite{jain2023two},
SolGPT~\cite{zeng2023solgpt}, \\
 DeFiAligner~\cite{gan2024defialigner}, RepairBench~\cite{silva2024repairbench}, 
 LLM-SmartAudit~\cite{wei2024llm}, \\Hyperion\cite{yang2024hyperion}, AdvSCanner~\cite{wu2024advscanner}, S{\'o}ley~\cite{soud2025soley}, Jiang et al.\cite{jiang2024unearthing}, \\ and XPLOGEN~\cite{eshghie2024highguard}
\end{tabular} \\  \midrule
Block Transaction Detection         & 3       & BLOCKGPT~\cite{gai2023blockchain}, Nicholls et al.~\cite{nicholls2023enhancing} and ZipZap~\cite{hu2024zipzap}   \\  \midrule
Contract Dynamic Analysis         & 3       & LLM4FUZZ~\cite{shou2024llm4fuzz}, ACFIX~\cite{zhang2024acfix} and  Sun et al.~\cite{sun2025adversarial}   \\  \midrule
Smart Contract Development        & 8       & \begin{tabular}[c]{@{}l@{}}Storhaug et al.~\cite{storhaug2023efficient}, karanjai et al.~\cite{karanjai2023smarter}, MazzumaGPT~\cite{dade2023optimizing},\\ Du et al.~\cite{du2024evaluation}, GPTutor~\cite{chen2023gptutor},
Petrovic et al.~\cite{petrovic2023model},\\ Zhao et al.~\cite{zhao2024automatic} and 
Haque et al.~\cite{haque2024extracting}
\end{tabular}   \\ \midrule
\begin{tabular}[c]{@{}l@{}}Cryptocurrency Community\\ Contributors\end{tabular}        & 5       & \begin{tabular}[c]{@{}l@{}}Trozze et al.~\cite{trozze2023large}, Axelsen et al.~\cite{axelsen2023scaling}, Liu et al.~\cite{liu2024decentralised},\\ ziegler et al.~\cite{ziegler2024classifying} and GPTutor~\cite{wahidur2024enhancing}\end{tabular}
   \\  \midrule
Miscellaneous        & 5       &  \begin{tabular}[c]{@{}l@{}}compilers~\cite{karanjai2024teaching}, zero-knowledge proofs~\cite{wellington2024basedai},\\ model training~\cite{luo2023bc4llm, gong2023dynamic} and NFT generation~\cite{he2023learning}\end{tabular}   \\ 
\bottomrule
\end{tabular}%
}
\end{table}

To fill the gap, we seek to delve into the multifaceted role of LLMs within the realm of blockchain security, exploring the comprehensive spectrum of LLMs on blockchain security (LLM4BS) tasks.
We commence by delineating the contemporary landscape of Large Language Model (LLM) applications across diverse domains (\S\ref{sec_overview_llm}), as well as the myriad of security threats implicated by the blockchain technology (\S\ref{sec_overview_blocksec}). 
Then, as illustrated in Table.\ref{tab_works}, we elaborate on the incorporation and progression of LLM4BS tasks, involving smart contract auditing, block transaction detection, contract dynamic analysis, smart contract development, and cryptocurrency community contributors (\S\ref{sec_llmbs}).
Thereafter, we meticulously select three quintessential cases of LLM4BS tasks to elucidate the state-of-the-art LLM4BS tasks (\S\ref{sec_cases}), comprising LLM4FUZZ~\cite{shou2024llm4fuzz}, SMARTINV~\cite{wang2024smartinv}, BLOCKGPT~\cite{gai2023blockchain}.
Finally, we present an insightful discourse on the challenges presently faced within the ambit of LLM4BS, and proffer prospective trajectories for future research and development in this emergent field (\S\ref{sec_future}).

This paper makes the following contributions:

\begin{itemize}

\item To the best of our knowledge, after a meticulous review of the existing literature, we conduct the first systematic examination focusing on the application of Large Language Models to tasks within the realm of blockchain security, offering a pioneering exploration of the interplay between advanced AI and distributed ledger systems. 

\item In our comprehensive survey, we meticulously chronicle the current landscape of applications of Large Language Models in the domain of blockchain security.
We delve into a detailed analysis of how Large Language Models are employed across various scenarios, from enhancing the reliability of smart contracts to fortifying the integrity of distributed ledger systems. This sheds light on the multifaceted contributions of this cutting-edge technology.

\item Based on our study, we rigorously compile and summarize a range of practical academic achievements related to the application of Large Language Models (LLMs) in strengthening blockchain security. We also propose several promising avenues for future research, anticipating that these will catalyze substantial advancements and innovations within this burgeoning intersection of fields.

\end{itemize}
\section{Overview of LLM4BS}
\label{sec_overview}

We provide basic knowledge about LLM4BS tasks in this section, including LLM applications in~\S\ref{sec_overview_llm} and threats of blockchain security in~\S\ref{sec_overview_blocksec}.

\subsection{Introduction to Large Language Models}
\label{sec_overview_llm}

This subsection will interpret the definition, characteristics, and diverse applications of Large Language Models (LLMs)

\subsubsection{Definition and Characteristics of LLMs}
\label{sec_overview_llm_definition}

Large Language Models (LLMs) represent a groundbreaking advancement in artificial intelligence, particularly within the domain of natural language processing (NLP)~\cite{hou2023large}. These models are characterized by their immense size, depth, and complexity, enabling them to process and generate human-like text with remarkable fluency and coherence~\cite{song2023llm}. At the heart of LLMs lies the transformer architecture, a powerful framework for sequence modeling that has revolutionized the field of NLP~\cite{zamfirescu2023johnny}.

The defining characteristics of LLMs include their unprecedented scale, which involves training on vast corpora of text data containing billions or even trillions of words. This extensive training data allows LLMs to capture the intricate nuances of language, including syntax, semantics, and pragmatics, thereby endowing them with a deep understanding of linguistic structures and conventions~\cite{kang2023large}. Additionally, LLMs exhibit a high degree of generative ability, capable of producing text that is contextually relevant and coherent across a wide range of tasks and domains.

Moreover, LLMs possess a remarkable degree of adaptability, thanks to their ability to be fine-tuned or specialized for specific applications or domains through techniques such as transfer learning~\cite{howard2018universal}. By leveraging pre-trained models and fine-tuning them on task-specific datasets, practitioners can tailor LLMs to address a diverse array of NLP tasks, ranging from sentiment analysis and language translation to document summarization and conversational agents~\cite{greshake2023not}.

Furthermore, LLMs demonstrate an advanced understanding of the context within language, enabling them to generate responses or predictions that are sensitive to the surrounding textual context~\cite{yin2020tabert}. This contextual awareness is achieved through mechanisms such as attention mechanisms and positional encodings, which enable LLMs to attend to relevant parts of the input sequence and model long-range dependencies effectively~\cite{dai2019transformer}.

Overall, LLMs represent a significant milestone in AI research and have unlocked new possibilities for human-computer interaction, content generation, information retrieval, and more. Their ability to understand and generate natural language at scale has led to transformative applications across various domains, shaping the future of AI-driven technologies~\cite{cheng2023compost}.

\subsubsection{Applications of LLMs in Various Domains}
\label{sec_overview_llm_apps}

\begin{figure}[tb]
	\centering
	\includegraphics[width=0.99\linewidth]{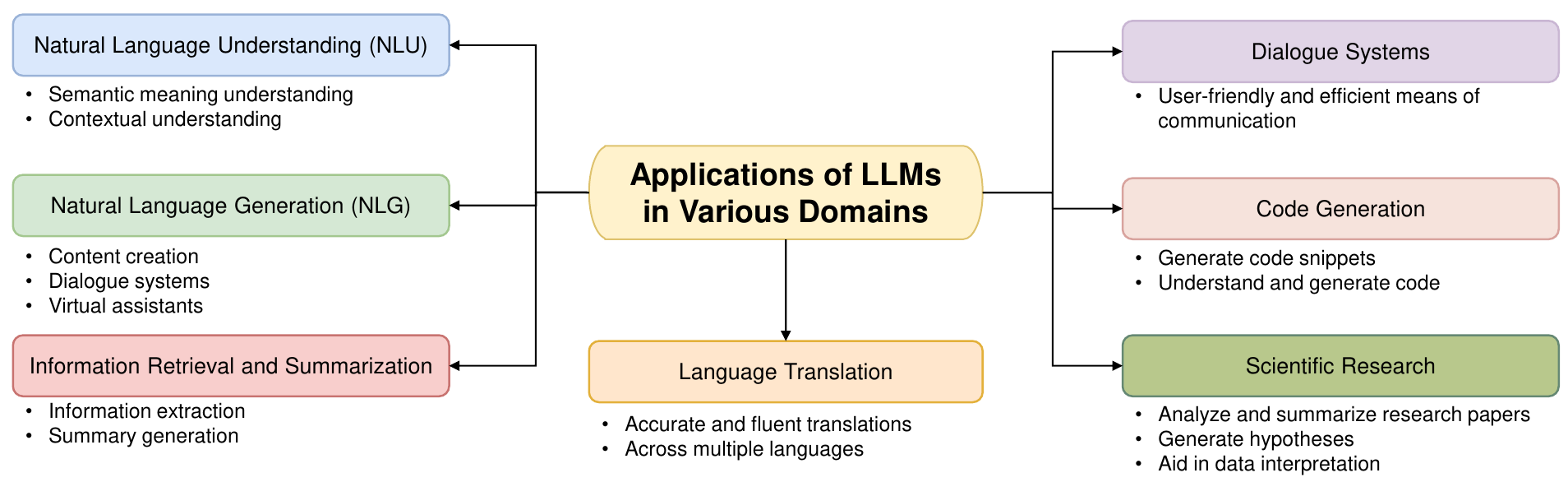}
	\caption{The various applications of LLM.}
	\label{fig_LLMApps}
\end{figure}

As depicted in Fig.~\ref{fig_LLMApps}, the versatility and efficacy of LLMs have led to their widespread adoption across diverse domains and applications, where they have demonstrated exceptional performance and utility~\cite{dai2023llm}. Some notable applications of LLMs include:

\textbf{Natural Language Understanding (NLU):} LLMs excel in tasks such as sentiment analysis, named entity recognition, and text classification, where the comprehension of semantic meaning and context is paramount~\cite{kim2023visually}. By leveraging their deep understanding of language, LLMs can accurately analyze and interpret textual data, enabling tasks such as sentiment analysis in social media monitoring or categorization of customer feedback.

\textbf{Natural Language Generation (NLG):} LLMs are proficient in generating human-like text for a variety of applications, including content creation, dialogue systems, and virtual assistants~\cite{ni2023lever}. Their ability to produce coherent and contextually relevant responses makes them invaluable for tasks such as generating product descriptions, composing personalized messages, or facilitating natural language interactions in conversational interfaces.

\textbf{Information Retrieval and Summarization:} LLMs play a crucial role in extracting relevant information from large volumes of text and generating concise summaries, thereby facilitating efficient information retrieval and knowledge extraction~\cite{mishra2023llm}. Whether summarizing news articles, extracting key insights from research papers, or generating abstracts for documents, LLMs offer a powerful solution for distilling vast amounts of textual data into digestible and informative summaries.

\textbf{Language Translation:} LLMs have revolutionized machine translation by providing more accurate and fluent translations across multiple languages~\cite{liu2022deplot}. By leveraging their vast linguistic knowledge and contextual understanding, LLMs can produce translations that preserve the meaning, tone, and style of the original text, enabling seamless communication across language barriers in various domains, including e-commerce, international diplomacy, and multicultural communication.

\textbf{Dialogue Systems:} LLMs power conversational agents and chatbots, enabling natural and contextually appropriate interactions with users~\cite{dong2023towards}. Whether assisting customers with product inquiries, providing personalized recommendations, or offering customer support, LLM-based dialogue systems offer a user-friendly and efficient means of communication, enhancing user experience and engagement.

\textbf{Code Generation:} LLMs are increasingly being used to generate code snippets and assist developers in programming tasks by understanding and generating code in various programming languages~\cite{ni2023lever,gu2023llm}. By analyzing code repositories and documentation, LLMs can generate code that adheres to programming conventions, syntax rules, and best practices, thereby accelerating the development process and aiding in code maintenance and debugging~\cite{vaithilingam2022expectation}

\textbf{Scientific Research:} LLMs support scientific discovery by analyzing and summarizing research papers, generating hypotheses, and aiding in data interpretation~\cite{agossah2023llm,dai2023llm}. By ingesting vast amounts of scientific literature and domain-specific knowledge, LLMs can assist researchers in navigating the ever-expanding body of scientific literature, identifying relevant publications, and extracting valuable insights to inform their research endeavors~\cite{sultanum2023datatales}.

These applications underscore the broad utility and transformative potential of LLMs across a wide range of domains and industries, highlighting their significance in advancing AI capabilities and enabling human-computer interaction at unprecedented levels of sophistication. As LLMs continue to evolve and improve, their impact on various fields is expected to grow, driving innovation, efficiency, and discovery in the years to come.

\subsection{Blockchain Security Fundamentals}
\label{sec_overview_blocksec}

This section will discuss the key components and common security threats of blockchain systems.

\subsubsection{Key Components of Blockchain Security}
\label{sec_overview_blocksec_keycomponents}

Blockchain security is a multifaceted endeavor aimed at safeguarding the integrity, confidentiality, and availability of data stored and processed within a blockchain network~\cite{he2024nurgle}. Key components of blockchain security include:

\textbf{Cryptography:} Cryptography lies at the heart of blockchain security, serving to encrypt data, authenticate participants, and ensure the integrity of transactions~\cite{kosba2016hawk,tan2021blockchain}. Techniques such as hashing, digital signatures, and cryptographic keys are utilized to secure data and verify the authenticity of transactions on the blockchain~\cite{dinh2018untangling}.

\textbf{Consensus Mechanisms:} Consensus mechanisms are protocols that govern how transactions are validated and added to the blockchain. By achieving agreement among network participants, consensus mechanisms ensure the immutability and integrity of the distributed ledger~\cite{sukhwani2017performance,li2020scalable}. Popular consensus mechanisms include Proof of Work (PoW)~\cite{gervais2016security}, Proof of Stake (PoS)~\cite{gavzi2019proof}, and Delegated Proof of Stake (DPoS)~\cite{li2017securing}, each with its own strengths and vulnerabilities.

\textbf{Decentralization:} Decentralization is a core principle of blockchain security, distributing control and decision-making authority across a network of nodes~\cite{conoscenti2017peer,monte2020scaling}. By eliminating single points of failure and reducing the risk of censorship or manipulation, decentralization enhances the resilience and security of the blockchain network~\cite{zyskind2015decentralizing}. However, achieving true decentralization requires careful consideration of factors such as node distribution, governance structures, and network incentives~\cite{li2018crowdbc}.

\textbf{Smart Contract Security:} Smart contracts are self-executing contracts with predefined rules and conditions encoded on the blockchain. Ensuring the security of smart contracts is essential to prevent vulnerabilities, exploits, and unauthorized access~\cite{zou2019smart,hu2020smart,chen2021sigrec,eshghie2021dynamic}. Techniques such as formal verification, code auditing, and secure development practices are employed to mitigate risks associated with smart contracts, including reentrancy attacks, integer overflow/underflow, and unchecked external calls~\cite{zhou2023sok,he2022tokencat,eshghie2024highguard}.

\subsubsection{Common Security Threats in Blockchain Systems}

\begin{figure}[tb]
	\centering
	\includegraphics[width=0.99\linewidth]{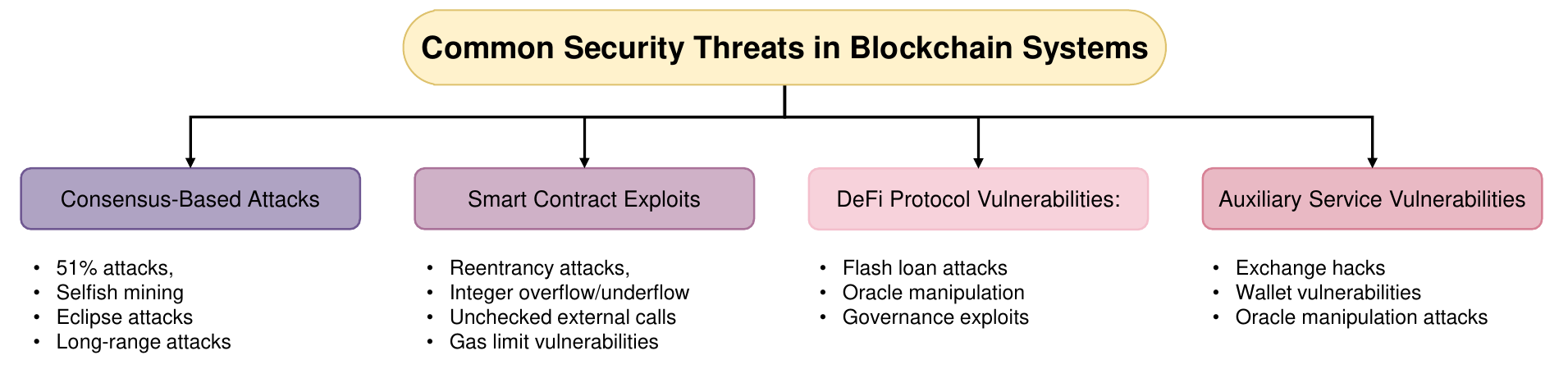}
	\caption{The threats in blockchain systems.}
	\label{fig_threats}
\end{figure}

Despite the robust security measures inherent in blockchain technology, various security threats and vulnerabilities pose risks to the integrity and functionality of blockchain systems~\cite{leng2020blockchain,berdik2021survey}.
We illustrate those threats in Fig.~\ref{fig_threats}.
Some common security threats in blockchain systems include:

\textbf{Consensus-Based Attacks:} Consensus-based attacks exploit vulnerabilities in the consensus mechanism to compromise the integrity or availability of the blockchain network~\cite{ma2020towards}. Examples include 51\% attacks, where a single entity or coalition controls the majority of the network's hash rate, enabling them to manipulate transaction confirmations or execute double spending attacks~\cite{xu2022sg}. Similarly, attacks such as selfish mining, eclipse attacks, and long-range attacks target weaknesses in specific consensus protocols, undermining the security and reliability of the blockchain network~\cite{xiao2020modeling}.

\textbf{Smart Contract Exploits:} Smart contract vulnerabilities pose significant risks to blockchain security, as they can be exploited to execute unauthorized transactions, drain funds, or trigger unintended behavior~\cite{brent2020ethainter,wan2021smart}. Common smart contract vulnerabilities include reentrancy attacks, where an attacker repeatedly calls a vulnerable contract's function before the previous invocation completes, enabling them to manipulate the contract's state and steal funds~\cite{sharma2023mixed,luo2024scvhunter,he2023tokenaware}. Other vulnerabilities, such as integer overflow/underflow, unchecked external calls, and gas limit vulnerabilities, can also be exploited to compromise the security of smart contracts and the underlying blockchain network~\cite{coblenz2019smarter,liao2022large}.

\textbf{DeFi Protocol Vulnerabilities:} Decentralized finance (DeFi) protocols introduce new security challenges due to their complex interactions and composability~\cite{chaliasos2024smart,zhou2023sok,chen2020soda}. Vulnerabilities in DeFi protocols, such as flash loan attacks, oracle manipulation, and governance exploits, can result in significant financial losses for users and undermine trust in the DeFi ecosystem~\cite{zhou2021just,wang2022impact,li2023demystifying}. Additionally, vulnerabilities in specific DeFi protocols can have cascading effects on other interconnected protocols, amplifying the impact of security breaches and systemic risks within the DeFi space~\cite{kong2023defitainter,gan2022understanding}.

\textbf{Auxiliary Service Vulnerabilities:} Auxiliary services, such as wallets, exchanges, oracles, and decentralized applications (DApps), serve as entry points for attackers to exploit vulnerabilities and compromise the security of blockchain systems~\cite{li2021strong,gan2023trick}. Security breaches in auxiliary services, such as exchange hacks, wallet vulnerabilities, or oracle manipulation attacks, can lead to the loss of funds, unauthorized access to user data, or manipulation of on-chain transactions~\cite{kim2023etherdiffer,li2021deter}. Furthermore, the interconnected nature of auxiliary services within the blockchain ecosystem amplifies the impact of security breaches, as vulnerabilities in one service can propagate to others, resulting in widespread disruption and financial losses.

Addressing these security threats and vulnerabilities requires a comprehensive approach that encompasses technical measures, best practices, and community collaboration to strengthen the resilience and security of blockchain systems~\cite{ma2020survey}. By understanding the key components of blockchain security and mitigating common security threats, stakeholders can foster greater trust, transparency, and adoption in the decentralized ecosystem, driving innovation and value creation for users worldwide.

\section{Taxonomy of LLM4BS tasks}
\label{sec_llmbs}

\begin{figure}[tb!]
	\centering
	\includegraphics[width=0.99\linewidth]{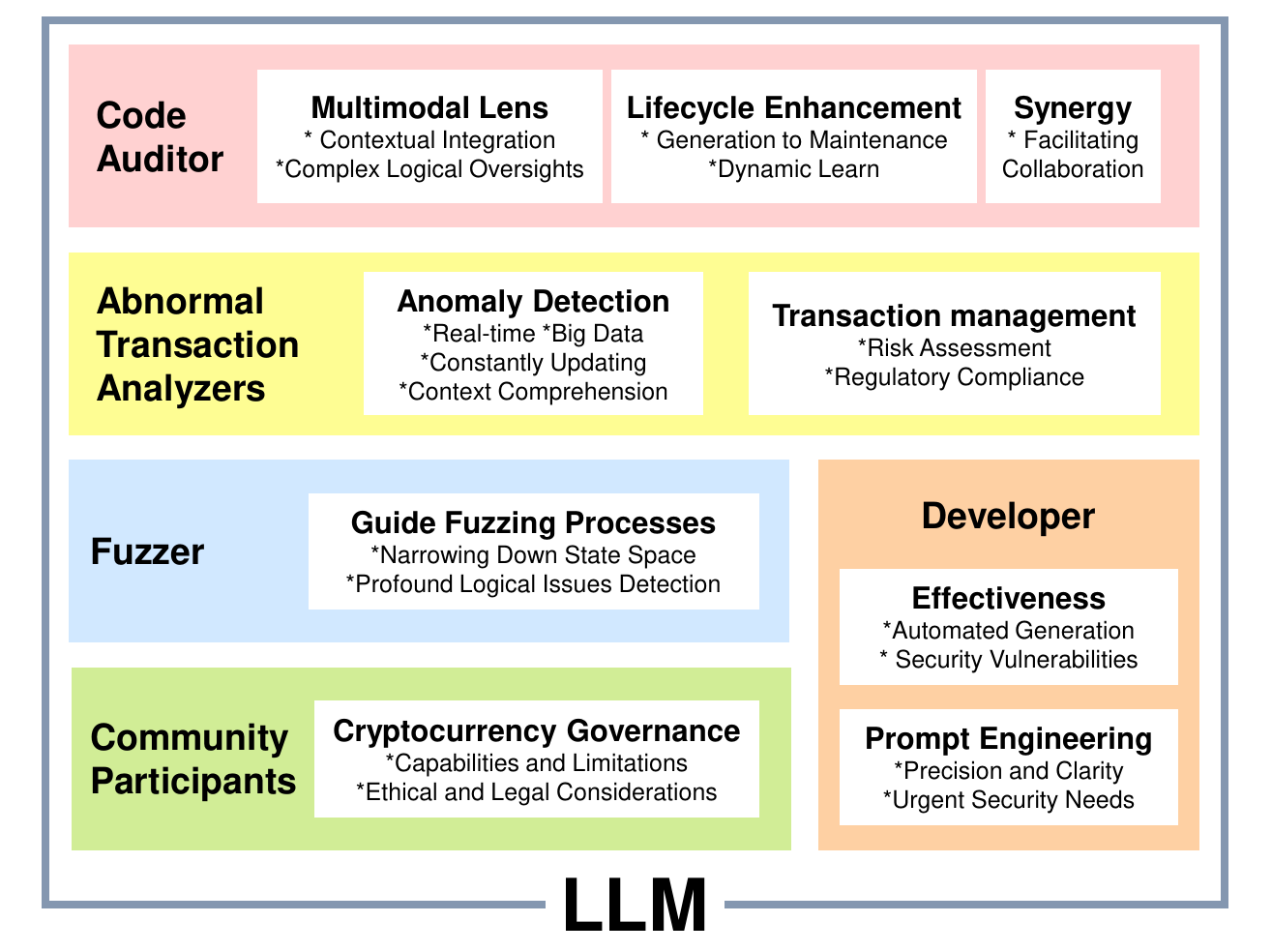}
	\caption{The applications of LLM on the task of blockchain security.}
	\label{fig_LLMsecApps}
\end{figure}

In this section, we introduce a thematic taxonomy devised to systematically categorize the body of literature about tasks associated with large language models for blockchain security (LLM4BS), emphasizing the function of the LLM within these contexts.
Fig.~\ref{fig_LLMsecApps} depicts the five applications of LLM4BS task, involving code audit of smart contracts~\S\ref{sec_llmbs_audit}, analysis of abnormal transactions~\S\ref{sec_llmbs_tx}, dynamic analysis of smart contracts~\S\ref{sec_llmbs_fuzz}, development of smart contracts~\S\ref{sec_llmbs_dev}, participants of cryptocurrency community~\S\ref{sec_llmbs_community}, and other potential directions~\S\ref{sec_llmbs_misc}.

\subsection{LLM as Code auditor on Smart Contracts}
\label{sec_llmbs_audit}

The application of LLM in the domain of smart contract code auditing and vulnerability detection can be succinctly encapsulated as follows: Advanced tools, such as SMARTINV~\cite{wang2024smartinv}, GPTScan~\cite{sun2024gptscan}, David et al.~\cite{david2023you}, Karanjai et al.~\cite{karanjai2023smarter}, ContractArmor~\cite{sonmez2024contractarmor}, Ortu et al.~\cite{ortu4530467identifying}, ASSBert~\cite{sun2023assbert}, PSCVFinder~\cite{yu2023pscvfinder}, LLM4Vuln~\cite{sun2024llm4vuln}, TrustLLM~\cite{ma2024combining},
AuditGPT~\cite{xia2024auditgpt}, 
Chen et al.~\cite{chen2023chatgpt}, Jain et al.~\cite{jain2023two},
 PropertyGPT~\cite{liu2024propertygpt}, SolGPT~\cite{zeng2023solgpt}, DeFiAligner~\cite{gan2024defialigner}, RepairBench~\cite{silva2024repairbench}, LLM-SmartAudit~\cite{wei2024llm}, Hyperion~\cite{yang2024hyperion}, AdvSCanner~\cite{wu2024advscanner}, S{\'o}ley~\cite{soud2025soley}, Jiang et al.\cite{jiang2024unearthing} and XPLOGEN~\cite{eshghie2024highguard}.
As shown in Table.\ref{tab_codeaudior},
these tools powered by Large Language Models signify a monumental shift from traditional, pattern-based analysis methodologies towards more contextually aware and comprehensive inspection techniques. These cutting-edge tools extend their analytical prowess beyond static patterns by knitting together disparate threads of information, including the nuanced aspects of natural language documentation that detail the intended functions and transactional constructs of smart contracts.
 
The integration of code and contextual data through a multimodal lens equips such tools with the capacity to unravel complex logical oversights and identify subtle "machine un-auditable" bugs, which would otherwise evade detection. By assimilating and interpreting the richer tapestry of human language explanations paired with code, LLM-based tools delve deeper into the intricate web of smart contract interactions. The profound understanding garnered from this approach not only sheds light on hidden vulnerabilities but also fortifies smart contracts against the myriad of risks that could lead to substantial financial repercussions.

In essence, the integration of Large Language Models in smart contract analysis marks a significant leap in safeguarding the infrastructural integrity of blockchain technology. It underscores an evolving landscape where artificial intelligence converges with software development practices to bolster security measures. This proactive identification and remediation of weaknesses within smart contracts, facilitated by the keen insights offered by LLMs, are instrumental in cementing trust and reliability in blockchain transactions—hence mitigating potential financial liabilities and reinforcing the bedrock of digital contracts.

\begin{table}[t!]
  \centering
\caption{Table of tools and models for code auditor}
\label{tab_codeaudior}
\resizebox{0.99\textwidth}{!}{%
\begin{tabular}{@{}lccr@{}}
\toprule
\textbf{Research}     & \textbf{Functionality}     & \textbf{LLM(s)}     & \textbf{Publications} \\ \midrule
SMARTINV~\cite{wang2024smartinv}        & Model finetuning approach           & Alpaca etc.  & SP,2024     \\
GPTScan~\cite{sun2024gptscan}         & Logic vulnerability detection       & GPT-3.5 etc. & ICSE,2024   \\
David et al.~\cite{david2023you}    & LLM audit feasibility               & GPT-4 etc.   & arXiv,2023       \\
Karanjai et al.~\cite{karanjai2023smarter} & LLM code evaluation                 & ChatGPT etc. & BRAINS,2023 \\
ContractArmor~\cite{sonmez2024contractarmor}   & Rule-based code analysis            & ChatGPT      & EUSPN,2024  \\
Ortu et al.~\cite{ortu4530467identifying}     & Contract automatic repair           & ChatGPT etc. & arXiv,2023       \\
Jain et al.~\cite{jain2023two}   & Contract automatic repair           & GPT-3.5 etc. & ICI,2023       \\
ASSBert~\cite{sun2023assbert}         & Contract vulnerability detection    & BERT         & JISA,2023   \\
PSCVFinder~\cite{yu2023pscvfinder}      & Contract vulnerability detection    & CodeT5       & ISSRE,2023  \\
Chen et al.~\cite{chen2023chatgpt} & Contract vulnerability detection & ChatGPT etc. & arXiv,2023\\
LLM4Vuln~\cite{sun2024llm4vuln}        & Vulnerability reasoning enhancement & GPT-4 etc.   & arXiv,2024       \\ 
TrustLLM~\cite{ma2024combining}        & Smart contracts audit & CodeLlama-13b   & arXiv,2024       \\ 
AuditGPT~\cite{xia2024auditgpt}        & ERC token audit & ChatGPT etc.  & arXiv,2024       \\ 
PropertyGPT~\cite{liu2024propertygpt}        & Formal verification automation & GPT-4  & arXiv,2024       \\ 
SolGPT~\cite{zeng2023solgpt}        & Contract vulnerability detection & GPT-2  & ICA3PP,2023       \\ 
DeFiAligner~\cite{gan2024defialigner}        & Logic inconsistency detection & GPT-3.5, GPT-4, GPT-4o  & AFT,2024       \\
RepairBench~\cite{silva2024repairbench}        & AI-driven program repair & GPT-4o etc. & arXiv,2024       \\ 
LLM-SmartAudit~\cite{wei2024llm} & Advanced vulnerability detection & GPT3.5, GPT-4o  & arXiv, 2024 \\
Hyperion~\cite{yang2024hyperion} & DApp inconsistency detection & LLaMA2 & arXiv, 2024\\
AdvSCanner~\cite{wu2024advscanner} & Reentrancy vulnerability exploitation & GPT-4o & ASE,2024\\
S{\'o}ley~\cite{soud2025soley} & Logic vulnerability detection & CodeBERT & JSS,2025
\\
Jiang et al.\cite{jiang2024unearthing} & LLM-based Code Smell Detector & GPT-4 & TSE,2024
\\
XPLOGEN~\cite{eshghie2024highguard} & Exploit Synthesis Tool & GPT-4 & ASE, 2024
\\
\bottomrule
\end{tabular}%
}
\end{table}

Expanding further on the key roles LLMs play, it's worth noting the vast potential these models have in enhancing the entire lifecycle of smart contract development~\cite{zhao2023deepinfer}. From generation to maintenance, LLMs facilitate the crafting of more secure and robust smart contracts. They do so by potentially providing recommendations during the development phase, suggesting best practices, and even generating code snippets that align with security guidelines. Throughout the auditing process, tools like GPTScan and SMARTINV can continuously learn and adapt to new patterns of vulnerabilities emerging from the evolving landscape of blockchain technology and cyber threats. This dynamic learning process is pivotal, as it allows for the development of increasingly refined models capable of detecting even the most covert and sophisticated vulnerabilities.

Moreover, the capacity of LLMs to assimilate context and understand code as it correlates to business logic makes them particularly effective in scenarios where contractual agreements are complex and layered with intricate logic. This is especially crucial in fields such as finance, where smart contracts govern transactions involving significant sums and numerous stakeholders. The vulnerability in such a domain could have catastrophic effects, not just financially but also in terms of reputational damage for the entities involved. Hence, the stakes in accurate and effective smart contract auditing cannot be overstated.

LLMs also enhance collaborative efforts throughout the industry by facilitating a common understanding among developers, auditors, and end-users. Their ability to parse and explain code in natural language bridges communication gaps, enabling stakeholders with varying levels of technical expertise to engage in meaningful dialogue regarding the security and functionality of smart contracts. This collaborative environment fosters a culture of shared responsibility and proactive engagement in addressing and preempting security concerns.

\subsection{LLM as Analyzers for abnormal transaction}
\label{sec_llmbs_tx}

\begin{table}[tb!]
  \centering
\caption{Table of tools and models for abnormal transaction analyzers}
\resizebox{0.9\textwidth}{!}{%
\begin{tabular}{@{}lccr@{}}
\toprule
\textbf{Research}     & \textbf{Functionality}     & \textbf{LLM(s)}     & \textbf{Publications} \\ \midrule
BLOCKGPT~\cite{gai2023blockchain}        & Transaction anomaly detection           & Transformer  & arXiv,2023     \\
Nicholls et al.~\cite{nicholls2023enhancing}         & Logic vulnerability detection       & BERT etc. & arXiv,2023   \\
ZipZap~\cite{hu2024zipzap}         & Unknown       & Unknown & WWW,2024   \\
\bottomrule
\end{tabular}%
}
\label{tab_txanalysis}
\end{table}

The application of LLMs for blockchain transaction analysis, such as BLOCKGPT~\cite{gai2023blockchain}, Nicholls et al.~\cite{nicholls2023enhancing} and ZipZap~\cite{hu2024zipzap}, underscores their crucial role in conducting real-time monitoring to detect signs of irregular or suspicious behavior. These tools in Table~\ref{tab_txanalysis} represent a significant advancement in the field, as they provide a more dynamic and adaptable approach to identifying potential threats within blockchain transactions.

Unlike static, rule-based systems, LLMs are capable of processing and learning from vast amounts of transaction data in real-time, which enables them to uncover not just known types of fraudulent activity, but also novel patterns that emerge as technology and attack methods evolve. By leveraging the power of machine learning, these models can constantly update their understanding of what constitutes normal transactional behavior. This continuous learning process is essential for adapting to the ever-changing landscape of blockchain technology and the complex strategies employed by malicious actors.

Furthermore, the adaptability of LLMs is not limited to pattern recognition—they also excel in understanding the context of transactions. This includes the analysis of smart contract interactions, execution traces, gas prices, and other transaction metadata that could provide hints about the legitimacy of a transaction. Contextual analysis allows LLMs to differentiate between legitimate, though unusual, transactional behavior and genuine anomalies that could indicate fraudulent activities, such as money laundering, phishing, or exploitation of contract vulnerabilities.

In addition to identifying potentially fraudulent transactions, LLMs also contribute to risk assessment and regulatory compliance. By analyzing the transaction data against current compliance standards and risk models, LLMs can assist financial institutions in managing their risk exposure and adhering to anti-money laundering (AML) and know-your-customer (KYC) regulations. Their sophisticated analysis capabilities can provide valuable insights to compliance officers and regulatory bodies, allowing for a more proactive approach to detecting and preventing financial crimes.

In summary, the application of LLMs in blockchain transaction analysis reflects a commitment to enhancing the security measures of digital financial systems. By combining deep learning algorithms with extensive transaction datasets, LLMs stand as a formidable line of defense, capable of not only identifying anomalous activities in real-time but also evolving with the advancing threats, ensuring a resilient and secure framework for managing blockchain-based transactions.

\subsection{LLM as Fuzzer for Smart Contract}
\label{sec_llmbs_fuzz}

\begin{table}[tb!]
  \centering
\caption{Table of tools and models for smart contract fuzzer}
\resizebox{0.8\textwidth}{!}{%
\begin{tabular}{@{}lccr@{}}
\toprule
\textbf{Research}     & \textbf{Functionality}    & \textbf{LLM(s)}     & \textbf{Publications} \\ \midrule
LLM4FUZZ~\cite{shou2024llm4fuzz}        & Fuzzing optimization tool           & Llama 2  & arXiv,2024     \\
ACFIX~\cite{zhang2024acfix}         & AC vulnerability repair       & GPT-4 & arXiv,2024   \\
Sun et al.~\cite{sun2025adversarial} & Adversarial LLM-based Fuzzing & GPT-4 & ASE,2025 \\
\bottomrule
\end{tabular}%
}
\label{tab_fuzzer}
\end{table}

Large Language Models (LLMs) have been increasingly employed to elevate the process of fuzzing, particularly in the realm of smart contract security analysis, such as LLM4FUZZ~\cite{shou2024llm4fuzz}, ACFIX~\cite{zhang2024acfix} and Sun et al.~\cite{sun2025adversarial}. This methodology in Table~\ref{tab_fuzzer} involves utilizing LLMs to accurately assess the complexity and vulnerability likelihood of specific code regions within a smart contract. Consequently, these metrics serve to guide the direction and focus of fuzzers, steering them toward code segments that are more likely to harbor potential security threats.
 
The application of LLMs to fuzzing exercises significantly elevates the efficiency of these operations by narrowing down the vast state space that fuzzers typically navigate. This precision-targeted fuzzing approach contributes to higher coverage and reveals more vulnerabilities than conventional tools, especially those pertaining to the intricate nature of smart contract code that traditional methods may overlook~\cite{wu2024we}.

Moreover, this refined fuzzing technique allows for the integration of user-defined invariants and manually inserted assertions to monitor and manage the state during fuzzing. This approach can reduce the exploration overhead and improve the detection of more profound logical issues that regular fuzzing routines might miss.
Evaluations of this LLM-enhanced fuzzing method within real-world decentralized finance (DeFi) projects have demonstrated its effectiveness, outperforming baseline fuzzing parameters and uncovering significant vulnerabilities. These vulnerabilities, if left undetected and exploited, could potentially result in substantial financial losses.

In summary, the fusion of LLMs into the fuzzing workflow offers a promising and intelligent solution to the challenges faced in automated security analysis of smart contracts, underscoring their potential for increasing the robustness of blockchain-based platforms.

\subsection{LLM as Developer for Smart Contract}
\label{sec_llmbs_dev}

\begin{table}[tb!]
\caption{Table of tools and models for smart contract developer}
  \centering
\label{tab_developer}
\resizebox{\textwidth}{!}{%
\begin{tabular}{@{}lccr@{}}
\toprule
\textbf{Research}     & \textbf{Functionality}     & \textbf{LLM(s)}     & \textbf{Publications} \\ \midrule
Storhaug et al.~\cite{storhaug2023efficient}        & Vulnerability-constrained decoding           & GPT-J-6B  & ISSRE,2023     \\
karanjai et al.~\cite{karanjai2023smarter}         & Smart contract generation       & ChatGPT etc. & BRAINS,2023   \\
MazzumaGPT~\cite{dade2023optimizing}        & Smart contract generation           & Davinci  & arXiv,2023     \\
Du et al.~\cite{du2024evaluation}         & Audit capacity evaluation       & GPT-4 & arXiv,2024   \\
GPTutor~\cite{chen2023gptutor}         & AI programming assistant       & GPT-3.5 etc. & arXiv,2023   \\
Petrovic et al.~\cite{petrovic2023model} & Smart contract generation & ChatGPT & ICSEng,2023 \\ 
Zhao et al.~\cite{zhao2024automatic} & AI programming assistant & GPT-3.5 &  arXiv,2024 \\
Haque et al.~\cite{haque2024extracting} & Norm extraction & ChatGPT & arXiv,2024 \\
\bottomrule
\end{tabular}%
}
\end{table}

Recent studies in Table~\ref{tab_developer}, such as Storhaug et al.~\cite{storhaug2023efficient}, karanjai et al.~\cite{karanjai2023smarter}, Petrovic et al.~\cite{petrovic2023model}, Zhao et al.~\cite{zhao2024automatic}, Haque et al.~\cite{haque2024extracting}, MazzumaGPT~\cite{dade2023optimizing}, Du et al.~\cite{du2024evaluation} and GPTutor~\cite{chen2023gptutor}, have begun to scrutinize the efficacy and reliability of Large Language Models (LLMs) like ChatGPT and Google Palm2 in the automated generation of smart contracts. These smart contracts are integral to the blockchain ecosystem, executing agreements without the need for intermediaries, and their accuracy and security are paramount. The research primarily constructs a testing framework that assesses smart contracts on multiple fronts, i.e., validity, correctness, efficiency, security, and maintainability.

These results have demonstrated that LLMs, despite showing proficiency in understanding contractual terms and generating syntactically correct Solidity code, often produce contracts with considerable security vulnerabilities. This finding signals a critical issue in the code's operational quality. The evaluations suggest that while LLMs can streamline the contract creation process, there's an underlying risk of generating code that could be exploited if used without a thorough review.

Importantly, the studies underscore the role of effective prompt engineering. It emerged that the LLMs' outputs are significantly influenced by the specificity and clarity of the prompts, which must be meticulously designed to minimize the risk of ambiguous or flawed code generation. This is particularly challenging because generating smart contracts requires precision, and the semantics of legal terms must be correctly interpreted and applied by the models.

These works point to the necessity for comprehensive analysis and improvement in the methodologies employed by LLMs. There is optimism that future iterations of LLMs, with better training and prompt design considerations, could enhance the quality and security of AI-generated smart contracts. It also hints at the potential for these tools to revolutionize contract generation by reducing the time and effort required, while flagging the urgent need for more robust security measures and testing methods.

Such research analysis provides an overarching view of the current state of LLM applications in smart contract generation. The discoveries made serve as a cautionary note about over-reliance on AI without adequate checks but also lay out a roadmap for future advancements that could harness AI's full potential responsibly.

\subsection{LLM as Participants for Cryptocurrency community}
\label{sec_llmbs_community}

\begin{table}[tb!]
\caption{Table of tools and models for Cryptocurrency community participants}
\centering
\label{tab_community}
\resizebox{\textwidth}{!}{%
\begin{tabular}{@{}lccr@{}}
\toprule
\textbf{Research}     & \textbf{Functionality}     & \textbf{LLM(s)}     & \textbf{Publications} \\ \midrule
Trozze et al.~\cite{trozze2023large}        &  Legal support tools           & GPT-3.5 etc.  & arXiv,2023     \\
Axelsen et al.~\cite{axelsen2023scaling}         & Community moderation support       & ChatGPT & arXiv,2023   \\
Liu et al.~\cite{liu2024decentralised}        & Blockchain-based Governance Framework           & \qquad \textbf{-} & IEEE Software,2024     \\
ziegler et al.~\cite{ziegler2024classifying}         & Automating Contextual Classification       & GPT-4 & arXiv,2024   \\
GPTutor~\cite{wahidur2024enhancing}         & Blockchain Revolutionizes Finance       & DistilBert etc. & IEEE Access,2024  \\
\bottomrule
\end{tabular}%
}
\end{table}

Large Language Models (LLMs) such as GPT-3.5 and ChatGPT are emerging as powerful tools in the cryptocurrency community, such as Trozze et al.~\cite{trozze2023large}, Axelsen et al.~\cite{axelsen2023scaling}, Liu et al.~\cite{liu2024decentralised}, Ziegler et al.~\cite{ziegler2024classifying} and GPTutor~\cite{wahidur2024enhancing}, albeit with their respective strengths and weaknesses. 
Related works in Table~\ref{tab_community} collectively depict a landscape where LLMs are being explored for their potential to revolutionize governance and legal processes within the high-stakes, highly volatile realm of cryptocurrency.

Governance emerges as a major theme, as LLMs could contribute significantly to the structuring and transparency of this largely unregulated space. The first document outlines the broader governance challenges faced by AI systems, suggesting blockchain as a viable solution to introduce verifiability and accountability. On the other hand, the limitations of LLMs in capturing the complexities of legal reasoning are highlighted, a concern that is echoed across the three studies to varying degrees.

The practical applications of these models in legal settings, specifically detailed in the second and third documents, emphasize their innovative role in drafting legal complaints. This development is promising for the future of legal work related to cryptocurrency regulations and litigation, as it suggests that LLMs could alleviate some of the workload from human experts, although the need for human oversight remains.

While governance and legal assistance dominate the discourse, there's a tone of cautious optimism throughout the texts. There is recognition of the transformative potential of LLMs in the cryptocurrency sector, but also a clear acknowledgment of the need for further advancement in AI technology to fully integrate into complex decision-making processes where legal and ethical considerations are paramount.

In essence, the collective narrative from the three documents converges on the premise that LLMs hold transformative potential for the cryptocurrency community's governance and legal sectors but must overcome challenges in understanding before they can be fully trusted in autonomous roles.

\subsection{Miscellaneous}
\label{sec_llmbs_misc}

\begin{table}[tb!]
\label{tab_others}
\caption{Table of other potential work}
\label{tab_others}
\resizebox{\textwidth}{!}{%
\begin{tabular}{@{}llll@{}}
\toprule
\textbf{Research}     & \textbf{Functionality}     & \textbf{LLM(s)}     & \textbf{Publications} \\ \midrule
SolMover~\cite{karanjai2024teaching}        &  Language Translation Framework           & Alpaca  & arXiv,2024     \\
BasedAI~\cite{wellington2024basedai}         & Privacy-Preserving Computation       & GPT-4 etc. & arXiv,2024   \\
BC4LLM~\cite{luo2023bc4llm}        & Secure Learning Path           & ChatGPT & arXiv,2023     \\
DLLM~\cite{gong2023dynamic}         & Dynamic Language Modeling       & \qquad \textbf{-} & arXiv,2023   \\
Diffusion-MVP~\cite{he2023learning}         & NFT creation platform       & Stable-Diffusion & MM,2023  \\
\bottomrule
\end{tabular}%
}
\end{table}

As displayed in Table~\ref{tab_others}, LLM is also used in other blockchain security fields, involving smart contract compilers~\cite{karanjai2024teaching}, zero-knowledge proofs~\cite{wellington2024basedai}, model training~\cite{luo2023bc4llm, gong2023dynamic}, NFT generation~\cite{he2023learning}. We will introduce their applications in detail in the future.
\section{Case study of LLM4BS}
\label{sec_cases}

\begin{table}[b!]
\centering
\caption{The table of the three cases on LLM4BS}
\resizebox{0.95\textwidth}{!}{%
\begin{tabular}{@{}lcccr@{}}
\toprule
Research & Domain               & Publications & Date & faculty                  \\ \midrule
LLM4FUZZ~\cite{shou2024llm4fuzz} & Fuzz                 & arXiv        & 2024 & UC Berkeley              \\ 
SMARTINV~\cite{wang2024smartinv} & Program analysis     & IEEE S\&P    & 2024 & Columbia University      \\ 
BLOCKGPT~\cite{gai2023blockchain} & Transaction analysis & arXiv        & 2023 & University of California \\ \bottomrule
\end{tabular}
}
\label{tab_3cases}
\end{table}

In this section, we engage in an in-depth examination through three distinct case studies, each serving to illustrate and shed light on the diverse and concrete applications of Large Language Models for Blockchain Systems (LLM4BS).
These cases in Table.\ref{tab_3cases} have been meticulously selected to encompass a broad range of scenarios, comprising LLM4FUZZ~\cite{shou2024llm4fuzz}~\S\ref{sec_cases_llm4fuzz}, SMARTINV~\cite{wang2024smartinv}~\S\ref{sec_cases_smartinv}, BLOCKGPT~\cite{gai2023blockchain}~\S\ref{sec_cases_blockgpt}.

\subsection{LLM4Fuzz}
\label{sec_cases_llm4fuzz}

\begin{figure}[t!]
	\centering
	\includegraphics[width=0.99\linewidth]{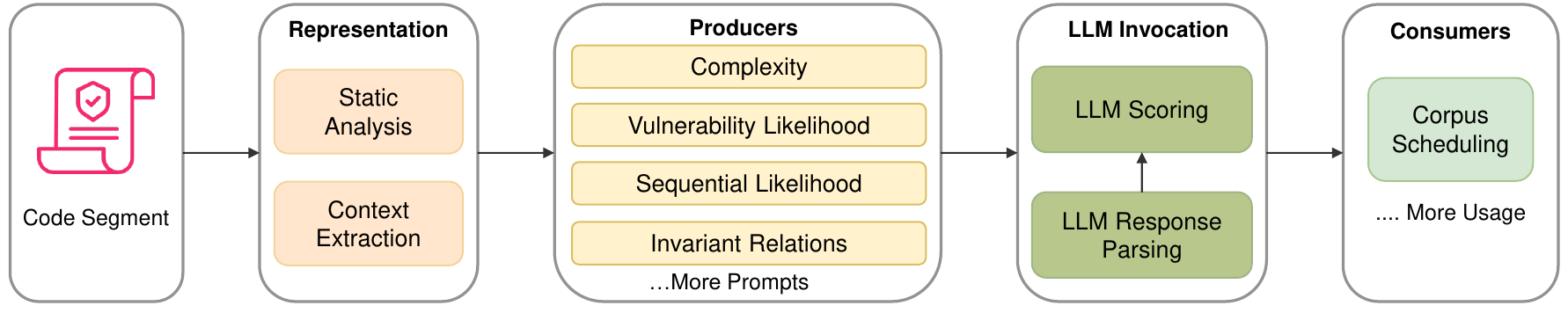}
	\caption{The architecture of LLM4FUZZ.}
	\label{fig_llm4fuzz}
\end{figure}

As depicted in Fig.\ref{fig_llm4fuzz}, LLM4FUZZ~\cite{shou2024llm4fuzz} emerges as an innovative technique in the cybersecurity landscape, specifically in the niche of smart contract security within blockchain networks. It intricately combines the prowess of Large Language Models (LLMs) with fuzz testing methodologies to proactively unearth vulnerabilities that could potentially compromise the integrity of smart contracts.

LLMs are highly sophisticated AI models that have made significant strides in understanding and generating human-like text, and more recently, they have proven to be adept at comprehending programming languages and code structure. LLM4FUZZ exploits this capacity by deploying LLMs to guide fuzzing processes intelligently. This results in a more incisive and nuanced exploration of smart contracts, focusing testing efforts on areas that LLMs determine to be most likely to contain security flaws. By doing so, LLM4FUZZ succeeds in not only streamlining the anomaly detection process but also in enhancing its accuracy and depth.

In the world of blockchain technology, where smart contracts serve as immutable agreements that execute automatically based on coded conditions, the potential negative impact of a security breach is heightened. Smart contracts control significant digital assets and are essential to the functioning of distributed applications (dApps). The immutable nature of blockchain adds a layer of complexity as deployed smart contracts, once committed to the blockchain, cannot be altered. Therefore, preemptive security assurances become crucial to ensuring their reliability and safeguarding the assets and processes they govern.

LLM4FUZZ provides a novel layer of security analysis by identifying and prioritizing potential problem areas within smart contract code. This prioritization is achieved through the LLM’s learned understanding of code patterns that are historically or commonly associated with vulnerabilities. The methodology enhances traditional fuzzing strategies, which typically adopt a more scattergun approach by bombarding the code with random data inputs. LLM4FUZZ’s targeted testing is not just more efficient but also more effective in discovering complex vulnerabilities that might otherwise be missed.

Following implementation, LLM4FUZZ has been benchmarked against existing fuzzing techniques and has consistently demonstrated superior performance. It expedites the vulnerability detection process and increases the breadth of security flaws that can be detected, thereby reinforcing the overall security posture.

The case of LLM4FUZZ is emblematic of the foresight in AI integration into cybersecurity regimes. It encapsulates the transformative effects of AI on improving and redefining existing technological processes, particularly in areas critical to the burgeoning digital economy. Through its lens, we catch a glimpse of the future of smart contract security – a future where AI-driven tools not only anticipate but actively engage in the continuous battle against cyber threats.

\subsection{SMARTINV}
\label{sec_cases_smartinv}

\begin{figure}[tb!]
	\centering
	\includegraphics[width=0.99\linewidth]{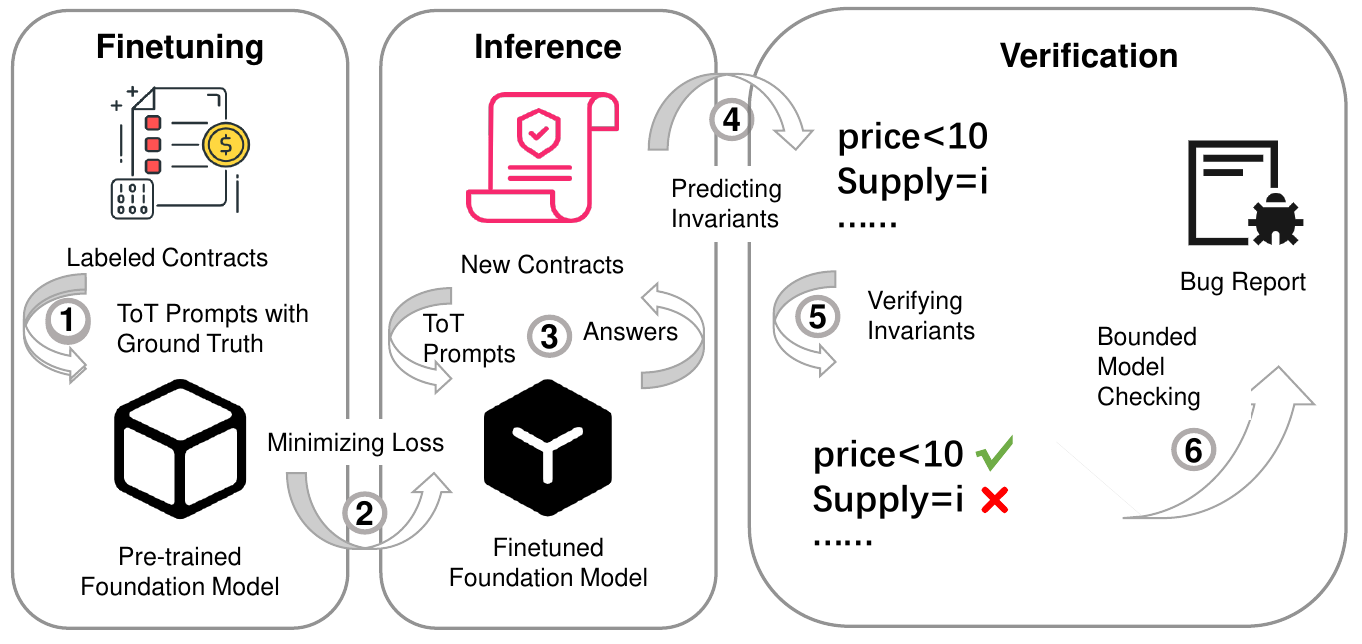}
	\caption{The architecture of SMARTINV.}
	\label{fig_smartinv}
\end{figure}

Proposed with the intention of enhancing the reliability and security of blockchain smart contracts, SMARTINV~\cite{wang2024smartinv} represents a significant breakthrough in the field. Its primary function is to infer invariants within smart contracts, which can be integral in automating the process of identifying elusive bugs that typically elude conventional machine-auditing methods. Fig.\ref{fig_smartinv} displays the architecture of SMARTINV.

The unique aspect of SMARTINV lies in its multimodal learning strategy, which acknowledges that truly understanding the operational behavior of smart contracts requires a multifaceted approach—one that combines and analyzes different types of information, or modalities. SMARTINV specifically leverages both the static code within a smart contract and dynamic transaction data. By correlating code patterns with transaction behaviors, SMARTINV is poised to uncover invariant conditions that point to a smart contract's expected and intended state throughout its lifecycle. This holistic approach ensures a more thorough examination and superior detection rate of potential security weaknesses that could lead to future vulnerabilities and exploits.

The framework operates on the premise that no singular mode of information can fully articulate a smart contract's intricate logic and potential edge cases. Hence, by fusing multiple data sources, SMARTINV captures a more accurate depiction of a smart contract's functionality, leading to a significant reduction in false positives and more precise bug detection. Such an integrated approach to smart contract analysis promotes greater assurance in their deployment and operation, which is a critical concern in blockchain applications where security and trust are paramount.

In deploying SMARTINV, the researchers demonstrate its efficacy by testing on a collection of smart contracts, where it shows not only a high degree of accuracy but also an impressive capability in scalability. SMARTINV emerges as an invaluable asset in the realm of smart contract development and auditing, setting a precedent for future methodologies to build upon its multimodal analysis framework for enhanced security measures in the ever-evolving domain of blockchain technology.

\subsection{BLOCKGPT}
\label{sec_cases_blockgpt}

\begin{figure}[tb!]
	\centering
	\includegraphics[width=0.99\linewidth]{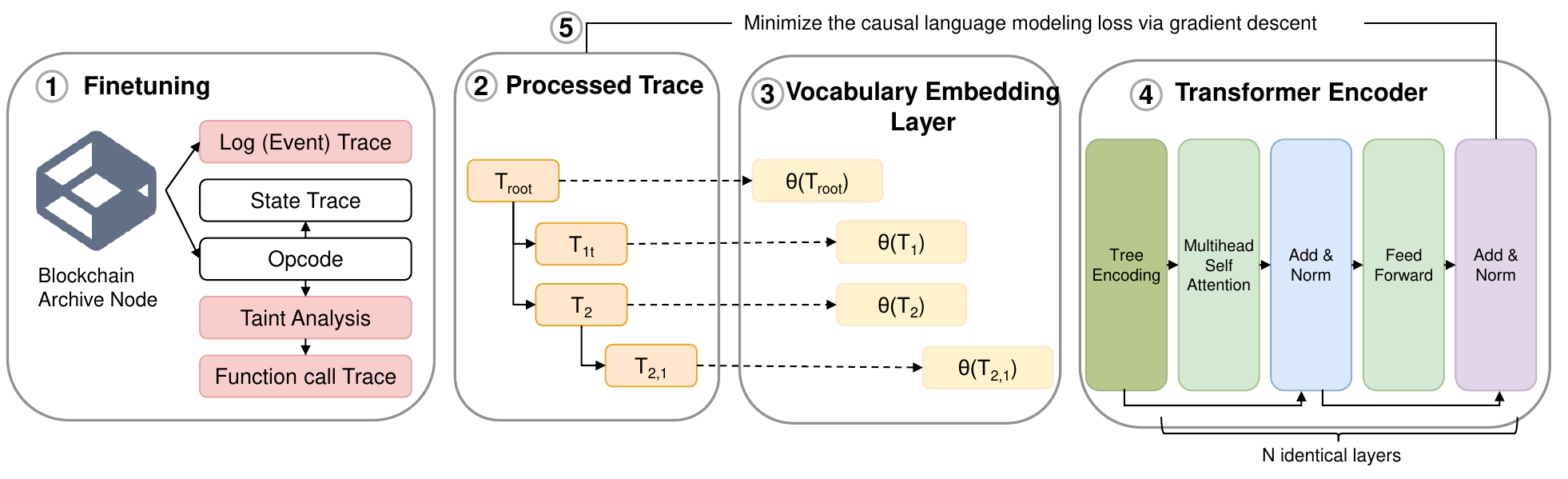}
	\caption{The architecture of BLOCKGPT.}
	\label{fig_blockgpt}
\end{figure}

As shown in Fig.\ref{fig_blockgpt}, BLOCKGPT~\cite{gai2023blockchain} serves as a paradigm shift in the domain of blockchain security, acting as a state-of-the-art Intrusion Detection System (IDS) specifically engineered to counteract and identify potentially malicious transactions within blockchain networks. The IDS is underpinned by a highly sophisticated large language model that has been meticulously trained with a significant corpus of transactional data from the Ethereum blockchain, one of the most widely utilized platforms in the industry.

The innovation expressed by BLOCKGPT is its departure from traditional detection methodologies that largely depend on predetermined rules or known patterns. Instead, BLOCKGPT adopts a proactive and learning-based approach that enables it to recognize a spectrum of anomalies, including sophisticated and previously unseen threats that could bypass conventional rule-based systems.

Demonstrating the prowess of its detection capabilities, BLOCKGPT has proven remarkably successful in testing scenarios. It proficiently identified and appropriately ranked 49 out of 124 verified attack transactions among the most abnormal three transactions that have occurred within their respective victim contracts. This high level of precision points to the system's refined anomaly recognition algorithms, indicating substantial progress in the field of IDS for blockchain.

Beyond its detection accuracy, the efficiency of BLOCKGPT is exemplified by its processing speed, handling transactions at an average rate of 2,284 per second, with relatively minimal deviation. This capability is not merely theoretical but is indicative of the system's readiness for deployment in real-world blockchain environments where real-time monitoring and response are critical.

The adaptability of BLOCKGPT extends to various blockchain architectures and applications, from finance to smart contracts. This versatility, combined with its real-time processing faculties, provides a robust and scalable solution that can be integrated seamlessly into existing blockchain infrastructures to fortify their resilience against a wide array of security threats.

As blockchain technology continues its integration into the fabric of digital transactions and smart contract deployment, systems such as BLOCKGPT represent vital components in the ongoing effort to safeguard these platforms. With the adoption of machine learning models like the one upon which BLOCKGPT is built, the future of blockchain IDS appears increasingly secure, paving the way for safer and more reliable blockchain operations.
\section{Future Direction and Challenge of LLM4BS tasks}
\label{sec_future}

In delving into the future of Large Language Models for Blockchain Security (LLM4BS), the academic community contends with a series of pivotal focus areas that necessitate concerted scholarly efforts to address inherent challenges and extend LLM's utility in blockchain systems. The following focal points are elaborated to reflect the nuances and complexity inherent in this field of study:

\textbf{Interdisciplinary Relationships:} The essence of the next stage in LLM4BS is undeniably grounded in a harmonized interplay among the domains of artificial intelligence, cyber protection mechanisms, and distributed ledger technologies~\cite{kim2023chatgpt,meyer2023chatgpt,motlagh2024large}. This interdisciplinary collaboration is not merely additive but synergistic, as it draws upon the strengths and insights of each discipline to forge a formidable shield against cyber animosities. There is a clarion call within the academic and industrial spheres for a robust alliance, emphasizing that the amalgamation of cognitive computing with cryptographic resilience and decentralized architectures can lead to a paradigm shift in securing blockchain networks.

\textbf{Regulatory and Compliance Challenges:} The shifting sands of regulatory frameworks demand not only compliance but a proactive engagement with regulatory bodies by scholars and practitioners in the LLM4BS field~\cite{wu2023brief,teubner2023welcome}. This relationship is reciprocal; as regulatory agencies develop a deeper understanding of the implications of integrating AI in blockchain, it is incumbent upon the actors within this space to advocate for regulations that encourage innovation while maintaining robust security measures. The dynamic interplay between cutting-edge technology and regulation is a delicate balance to strike, fostering a stable yet flexible platform for growth and adaptation in blockchain security solutions.

\textbf{Dynamic Security Threats:} The cyber threat horizon is akin to a chimeric beast—constantly mutating and presenting unforeseen challenges~\cite{liu2023summary,tan2023can}. Security models like LLM4BS must be engineered with inherent plasticity, allowing them to evolve alongside the threats they are designed to counteract. The integration of LLMs in blockchain security is not a static solution but a continually adapting safeguard, necessitating an expansive approach to cybersecurity that accounts for the proliferation of sophisticated cyberattacks as well as the subtleties of targeted breaches. Sustaining the integrity of blockchain transactions hinges on the preemptive identification and neutralization of these mercurial threats.

\textbf{Ethical Governance and Bias Mitigation:} The ethical tapestry within which LLM4BS operates is rich and complex, mandating a conscientious approach towards the examination and resolution of security practices that may inadvertently propagate bias or unfair outcomes~\cite{aydin2023chatgpt,ray2023chatgpt}. The quest for equitable algorithms expands beyond the technical realm, engaging with sociocultural dynamics and the moral dimensions of technological deployments. Therefore, a concerted effort in research that transcends statistical bias mitigation, touching upon philosophy, sociology, and ethics, is essential for fostering a climate where AI not only fortifies security but does so with an underlying commitment to justice and fairness.

\textbf{Energy Considerations and AI Sustainability:} In addressing the carbon footprint of blockchain operations, there is also a pressing need to confront the energy-intensive nature of training and deploying Large Language Models~\cite{roumeliotis2023chatgpt,qin2023toolllm,miao2023dao}. The ecological impact of these AI systems necessitates a dual strategy: enhancing algorithmic efficiency to reduce computational load and exploring alternative energy sources that can power these activities sustainably. This pursuit of ecological harmonization in the application of LLM4BS must be reflective of a broader commitment to sustainability across all aspects of blockchain technology, ensuring that the acceleration of security capabilities does not come at an unsustainable environmental cost.


\textbf{Ethical Considerations in AI}: The role of ethics cannot be overstated in the trajectory of LLM4BS implementation, as it undergirds every facet of AI application—from the source of data to the transparency of algorithms and the accountability for decisions made by or with the aid of AI. Implementing a robust ethical framework for LLM4BS entails a deep interrogation of the principles guiding AI development, encouraging scrutiny that permeates every layer of model design, deployment, and monitoring. Thus, creating an environment where trust in AI-fueled security measures is not merely assumed but carefully cultivated through responsible practices.

\textbf{Data Quality and Access}: At the heart of robust LLM4BS deployments lies the foundational element of data—its caliber, its scope, and the accessibility afforded to it. Herein lies the challenge: constructing and maintaining databases that are not only comprehensive and representative but are also curated with an eye toward enhancing the efficacy of Large Language Models in detecting anomalies and reinforcing security parameters in blockchain transactions. The task extends to crafting protocols that ensure data integrity and sourcing that conforms to ethical standards, thereby upholding the sanctity and reliability of these AI systems.

Navigating these considerations requires a strategic, methodological approach to utilize the full promise of LLM4BS. This involves a commitment to ongoing research, rigorous ethical scrutiny, and a concerted effort to evolve in tandem with the technological and regulatory landscape. With a fundamental understanding of these points, the community is better equipped to pave the way for LLM4BS to enhance the resilience and efficiency of blockchain security measures.
\section{Conclusion}
\label{sec:sample1}
In conclusion, our review of the integration of Large Language Models (LLMs) into blockchain security highlights the technological advancements and intricate challenges presented by this combination of LLM4BS. The potential of LLMs to enhance security protocols in the blockchain is evident, offering innovative solutions for smart contracts, abnormal transaction detection, and cryptocurrency community development. However, realizing this potential requires vigilance regarding scalability, privacy, evolving cyber threats, and the ethical implications of AI. 
The success of LLMs in blockchain security hinges not only on continuous technological refinement but also on ethical practices, regulatory alignment, and informed community engagement. The integration of LLMs into blockchain security marks a transformative era that necessitates a collaborative approach, balancing innovation with prudent oversight to forge a resilient and equitable security future.



\bibliographystyle{elsarticle-num-names} 
\bibliography{ref}





\end{document}